\documentclass[twocolumn,10pt,prl, nofootinbib]{revtex4}

\usepackage[dvips]{graphicx}
\usepackage{epsfig,amsmath,amssymb,verbatim,mathrsfs,array,layout,textcomp,amssymb,latexsym,slashed,graphicx,hyperref,booktabs,color,cleveref,mathtools,tikzfeynman}

\renewcommand*{\arraystretch}{1.5}

\newcommand{\beq}{\begin{equation}}
\newcommand{\eeq}{\end{equation}}
\newcommand{\bea}{\begin{eqnarray}}
\newcommand{\eea}{\end{eqnarray} }
\newcommand{\no}{\nonumber}

\newcommand{\del}{\partial}
\newcommand{\hf}{\frac{1}{2}}

\begin{document}
\title{Viable Twin Cosmology from Neutrino Mixing}

\author{Csaba Cs\'aki}\email{csaki@cornell.edu}
\author{Eric Kuflik}\email{kuflik@cornell.edu}
\author{Salvator Lombardo}\email{sdl88@cornell.edu}

\affiliation{Laboratory for Elementary Particle Physics, Cornell University,
Ithaca, NY 14850, USA}

\begin{abstract}

Twin Higgs models solve the little hierarchy problem without introducing new colored particles, however they are often in tension with measurements of
the radiation density at late times.
Here we explore viable cosmological histories for Twin Higgs models. In particular, we show that mixing between the SM and twin neutrinos can
thermalize the two sectors below the twin QCD phase transition, significantly reducing the twin sector's contribution to the radiation density.
The requisite twin neutrino masses of ${\mathcal{O}(1-20)}$ GeV and mixing angle with  SM neutrinos  of $10^{-3} - 10^{-5}$ can be probed in a
variety of current and planned experiments. We further find that these parameters can be naturally accessed in a warped UV completion, where the
neutrino sector can also generate the $Z_2$-breaking Higgs mass term needed to produce the
   hierarchy between the symmetry breaking scales $f$ and $v$.
\end{abstract}

\maketitle

\section{Introduction}

Twin Higgs (TH) models provide an elegant solution to the hierarchy problem without introducing new states that are charged under the SM gauge
symmetries~\cite{Chacko:2005pe}. Instead, a mirror sector with its own SU(3)$\times$SU(2)$\times$U(1) gauge symmetry is assumed. The $Z_2$ symmetry relating the SM and mirror sectors protects the Higgs mass from large radiative corrections, with the twin partners cancelling the
SM quadratic divergences at one loop. Other variations of neutral naturalness include~\cite{Burdman:2006tz,Cai:2008au,Craig:2014aea,Craig:2014roa,Chacko:2005vw,Batell:2015aha,Arkani-Hamed:2016rle}. While this idea is very efficient at hiding new
physics from the LHC and future colliders, it often leads to tension with cosmological observations due to the appearance of new light relativistic
degrees of freedom (DOF), namely the twin photon and twin neutrinos.

The standard assumption of TH models is that only the Higgs portal connects the SM and the mirror sectors. This maintains thermal
equilibrium between the two sectors down to temperatures of a few GeV, below which the twin sector decouples~\cite{Barbieri:2005ri}. At this point the twin and SM sectors have similar energy densities, and the twin photon and neutrinos contribute significantly to the radiation density at late times.  In particular, the Mirror Twin Higgs (MTH) model---the scenario where the mirror sector is a full copy of the SM---predicts an exceedingly large
contribution to the correction of the total radiation density (usually expressed in terms of $\Delta N_{\rm eff}$, as measured from Big Bang
Nucleosynthesis (BBN) and the Cosmic Microwave Background (CMB)).

Recently, several solutions have been proposed for the cosmological problems of the Twin Higgs, including the Fraternal Twin Higgs (FTH)~\cite{Craig:2015pha}, hard $Z_2$-breaking in the Yukawa couplings~\cite{Barbieri:2016zxn}, and SM reheating from a light right-handed neutrino~\cite{Chacko:2016hvu}. Further  cosmological aspects of Twin Higgs models, including dark matter, have been studied in~\cite{Craig:2016lyx, Chacko:2016hvu, Barbieri:2016zxn,
Prilepina:2016rlq,Farina:2015uea,Farina:2016ndq,Garcia:2015toa,Craig:2015xla,Garcia:2015loa}.

In this paper we propose that the neutrino portal can also naturally be used to connect the twin and the SM sectors. Mixing between the SM and
twin neutrinos appears in many simple implementations of the twin neutrino sector. We show that such mixing can lower the decoupling temperature
between the two sectors, potentially delaying decoupling past the scale of the twin QCD phase transition. 
When the decoupling of the two sectors happens
between the twin and ordinary QCD phase transition scales, the contribution of the twin sector to $\Delta N_{\rm eff}$ is strongly reduced,
since at the time of equilibrium there are fewer relativistic DOFs in the twin sector compared to the SM.  We explore in detail the
dependence of $\Delta N_{\rm eff}$ on the decoupling temperature, as well as the relation between the decoupling temperature and the twin neutrino masses and mixings with the SM neutrinos. We find that reasonably sized mixing between the two sectors of order $\sin \theta \sim 10^{-3}-10^{-5}$, and twin neutrino masses of $\mathcal{O}(10)$~\rm GeV, can  result in a viable cosmological
scenario for TH models.

We also show that a composite Twin Higgs
(CTH) UV completion~\cite{Geller:2014kta, Csaki:2015gfd, Batra:2008jy, Barbieri:2015lqa, Low:2015nqa} can naturally incorporate twin neutrino masses and mixings with the SM neutrinos of the desired magnitude.  By using this CTH framework we demonstrate that
the neutrino sector can also automatically generate the $Z_2$-breaking Higgs mass term needed to produce the hierarchy between the symmetry breaking scales $f$
and $v$ for the same parameters that result in a viable cosmology. 

This paper is organized as follows: We first investigate the dependence $N_{\rm eff}$ on the decoupling temperature of the twin sector. Then we calculate the decoupling temperature as a function of the twin neutrino masses and their mixings with the SM neutrinos, followed by an overview of the possible new experimental signals of the various TH scenarios. Next we present realistic mass and mixing patterns in the neutrino sector, followed by an implementation of this sector in the holographic CTH setup. We close the paper by a discussion of the $Z_2$-breaking effects in the Higgs potential. Various appendices contain the details of the RS construction,  the resulting warped mass spectrum, the effect of Majorana masses on the spectral functions, and finally the details of the full Coleman-Weinberg calculation for the neutrino sector.

\section{Dark radiation in Twin Higgs models\label{sec:Neff}}

We begin by describing the contributions to the radiation density of the universe at late times in various types of Twin Higgs models and compare those to the experimental bounds. Later we will show how to use mixing in the neutrino sector to obtain viable scenarios.

 The total radiation density of the universe is typically parameterized in terms of the effective number of neutrino species, $N_{\rm eff}$, defined as
\beq
\rho _{r} \equiv \rho_{\gamma} \left( 1+ \frac{7}{8} \left( \frac{4}{11}\right)^{4/3} N_{\rm eff}\right)\,,
\eeq
where $\rho_\gamma$ is the observed radiation density and $N^{\rm SM}_{\rm eff}=3.046$ is  the value predicted in the SM from standard neutrino decoupling.

For a twin sector identical to the SM, the total energy density of the universe doubles,
leading to $\Delta N_{\text{eff}} \equiv N_{\rm eff} -N_{\rm eff}^{\rm SM} \simeq 7.4$, which is very strongly excluded by the Planck result of $N_{\text{eff}} = 3.15 \pm 0.23$~\cite{Ade:2015xua}. Of course, we already know that the twin sector cannot be identical to the SM sector, since at the very least the Higgs vev ratios obey  $f/v > 1$. The simplest solution to avoid the $N_{\rm eff}$ constraint would be to raise the mass of all the light twin particles, which would remove the twin contributions to $N_{\text{eff}}$. This however is not possible: the twin electron (or the twin tau for the case of the FTH) would not be able to annihilate away and would overclose the universe. Therefore at least one of the twin states must remain light to allow the annihilation of the twin electrons.

Since there are necessarily contributions to $\Delta N_{\rm eff}$, we need to refine the prediction by taking into account the temperature difference between the two sectors: the value of $N_{\rm eff}$ will be determined by $g_\star'$,
\beq
N_{\rm eff} = N_{\rm eff}^{\rm SM} + \frac{4}{7} \left( 11/4 \right)^{4/3} g_\star^\prime
\eeq
where
\beq
g_\star^\prime = \sum_i s_i  g_i  \left( \frac{T_i^\prime}{T} \right)^4
\eeq
is the number of effective degrees of freedom in the twin sector weighted by the relative temperatures of each component  $T^\prime_i$ compared to the SM, $s_i = 1 (7/8)$ for a boson (fermion), and the sum runs over the relativistic twin DOFs at late times. The ratio of the temperature of the dark sector to the SM temperature can be calculated assuming separate entropy conservation in both  sectors after decoupling \cite{Feng:2008mu},
\beq
\frac{{T^\prime}}{T} = \left(\frac{g_{\star s}(T)}{{g}^\prime_{\star s}(T)} \frac{{g}^\prime_{\star s}(T_d)}{{g}_{\star s}(T_d)} \right)^{1/3} ~~~~~~~~~~{\rm for~} T< T_d
\eeq
where $T_d$ is the temperature at decoupling. Next we will  consider various possible options for lowering $\Delta N_{\rm eff}$ in different realizations of Twin Higgs models.

The simplest approach to lowering $\Delta N_{\rm eff}$ in Twin Higgs models is to raise the mass of the twin neutrinos above a few GeV.  This will remove the twin neutrino's contributions to $g_\star^\prime (T_d)$ and to $g_\star^\prime (T)$, resulting in a smaller $\Delta N_{\text{eff}}$, where the additional energy density at late times arises entirely from the twin photon. This can be naturally achieved by lowering the twin seesaw scale, but comes at a price of an additional source for $Z_2$ breaking. In a later section, we show that this could potentially also be the source of the $Z_2$ breaking in the Higgs potential generating $f/v \sim$ a few.

\begin{figure}[t!]
\center{
\includegraphics[width=.47\textwidth]{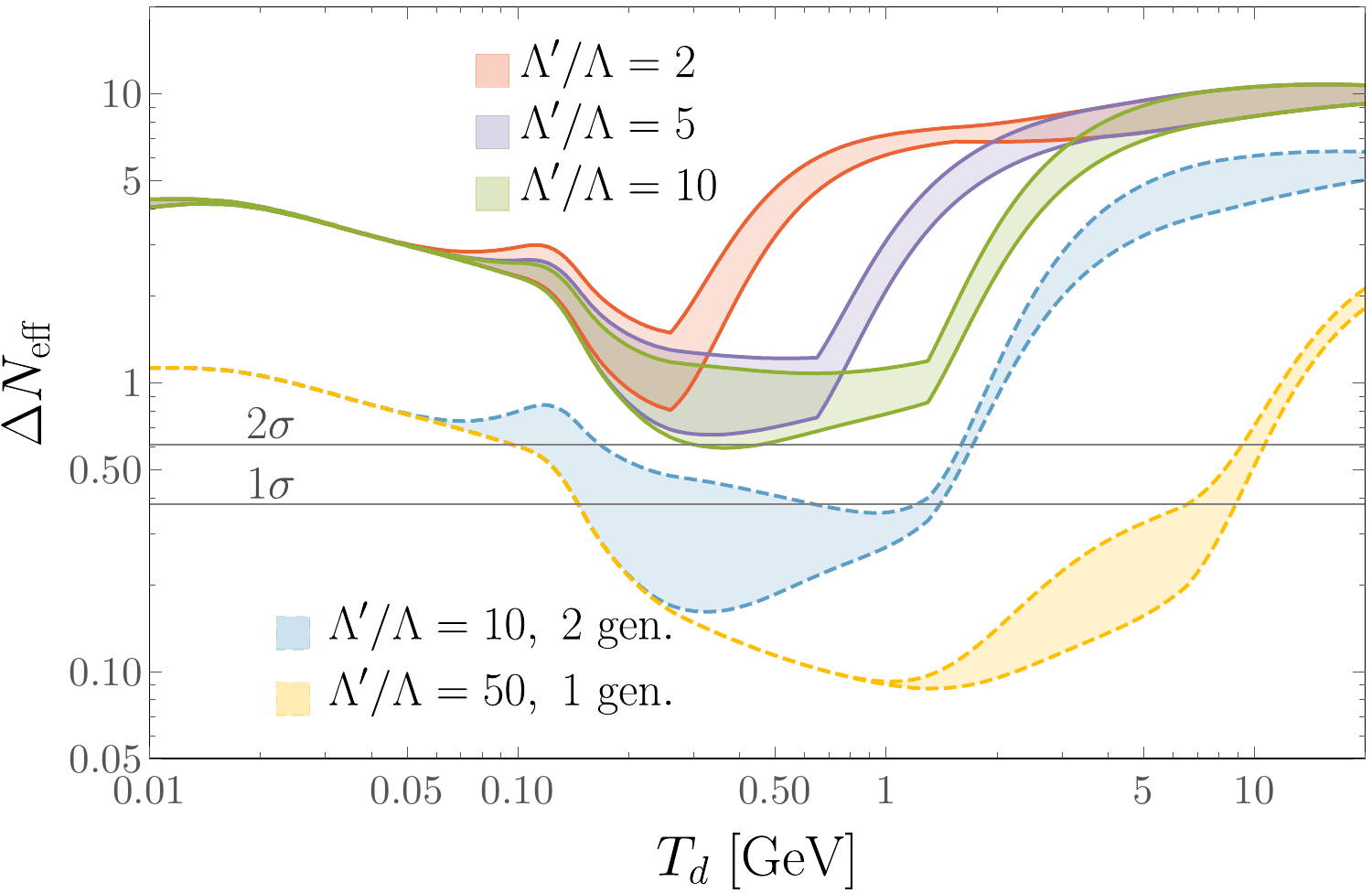}
\caption{Contribution to $\Delta N_{\rm eff}$ with heavy twin neutrinos ($m_{\tilde{\nu}}=10$ GeV) and massless twin photon, when varying $\Lambda^\prime/\Lambda$. Each band corresponds to varying $f/v$ from 5 (top of band) to 20 (bottom of band). The smallest contribution is obtained when the two sectors decouple between their QCD phase transitions. The dashed bounded regions correspond to one and two generation Twin Higgs models. The Planck constraint of $N_{\text{eff}} = 3.15 \pm 0.23$~\cite{Ade:2015xua} is also shown.
 \label{NeffMTH}}}
\end{figure}

The resulting predictions for $\Delta N_{\rm eff}$ depend significantly on the value of the decoupling temperature, $T_d$. This a consequence of the large change in the number of degrees of freedom during the twin (and ordinary) QCD phase transition (PT). If decoupling happens between the two PTs, then $\Delta N_{\rm eff}$ will be strongly reduced.
The twin QCD PT strongly reduces the degrees of freedom within the twin sector, while dumping its entropy into both sectors as they are still in equilibrium. 
Then, if by the time the SM QCD PT occurs, the two sectors have thermally decoupled, the entropy of the SM QCD PT will be  transferred only to the SM bath, raising the SM temperature relative to the twin temperature.
This suppression will be reflected in the final value of $\Delta N_{\rm eff}$ measured at late times.


In Fig.~\ref{NeffMTH} we show the predictions for $\Delta N_{\text{eff}}$ for the  MTH model with a massless twin photon, and with the masses of the three twin neutrinos  raised to $m_{\tilde{\nu}} = 10$ GeV, as a function of the decoupling temperature $T_d$. Here $f/v$ has been varied from 5 to 20, and the ratio of twin to SM QCD phase transition temperatures, $\Lambda^\prime / \Lambda$, between 2 and 10 (where $\Lambda^\prime/\Lambda =5$ is obtained in MTH with an $\mathcal{O}(10\%)$ splitting of the SU(3) gauge couplings~\cite{Farina:2015uea}). The minimum contribution to $\Delta N_{\text{eff}}$ depends significantly on $f/v$ for any decoupling temperature, as it determines whether or not the light twin states---namely the twin pions, muons and electrons---are relativistic at the time of decoupling. There is no strong dependence of the minimum contribution to $\Delta N_{\text{eff}} $ on $\Lambda^\prime/\Lambda$; however, for $\Lambda^\prime/\Lambda \sim 1$,  the value of $T_d$ required to avoid the constraint has to lie in a narrow range between the two QCD PT scales.

We learn that without introducing an additional source of $Z_2$ breaking in the twin Yukawa couplings, MTH is in tension with the Planck constraint on $\Delta N_{\rm eff}$.  In order to satisfy the 2$\sigma$ bounds of $\Delta N_{\text{eff}} = 0.61$, $f/v \gtrsim 20$ is required, which would imply reintroducing tuning into the Higgs potential. Additionally, we show $N_{\rm eff}$ for Twin Higgs models with one or two generations. The one generation model can be thought of as a FTH model with gauged hypercharge. Stage 3/4 CMB experiments should be able to highly constrain Twin Higgs models with a single light state, either a twin photon or neutrino, independent of $T_d$, $f/v$, and $\Lambda^\prime / \Lambda$, since such scenarios contribute a minimum $\Delta N_{\text{eff}} \gtrsim $ 0.088 or 0.065, respectively.

\section{Thermal decoupling from neutrino mixing \label{sec:decoupling}}

\begin{figure}[t!]
\begin{center}
\begin{tikzpicture}[line width=1.5 pt, scale=1]
\hspace{.3cm}
	\draw[fermionnoarrow] (-135:1)--(0,0);
	\draw[fermionnoarrow] (135:1)--(0,0);
	\draw[vector] (0:1)--(0,0);
	\node at (-135:1.2) {$\nu$};
	\node at (135:1.2) {$\tilde{\nu}$};
	\node at (.5,.3) {$Z$};	
	
	\begin{scope}[rotate=135]
		\begin{scope}[shift={(.5,0)}] 
			\clip (0,0) circle (.175cm);
			\draw[fermionnoarrow] (-1,1) -- (1,-1);
			\draw[fermionnoarrow] (1,1) -- (-1,-1);
		\end{scope}	
	\end{scope}

\begin{scope}[shift={(1,0)}]
	\draw[fermionnoarrow] (-45:1)--(0,0);
	\draw[fermionnoarrow] (45:1)--(0,0);
	\node at (-45:1.2) {$\nu$};
	\node at (45:1.2) {$\nu$};	
\end{scope}

\hspace{-.6cm}
\begin{scope}[shift={(5,-.3)}]
	\draw[fermionnoarrow] (-30:1)--(0,0);
	\draw[fermionnoarrow] (180:1)--(0,0);
	\draw[vector] (30:1)--(0,0);
	\node at (-30:1.2) {$\nu$};
	\node at (180:1.2) {$\tilde{\nu}$};
	\node at (.3,.6) {$Z$};	
	
		\begin{scope}[rotate=0]
		\begin{scope}[shift={(-.5,0)}] 
			\clip (0,0) circle (.175cm);
			\draw[fermionnoarrow] (-1,1) -- (1,-1);
			\draw[fermionnoarrow] (1,1) -- (-1,-1);
		\end{scope}	
	\end{scope}

\begin{scope}[shift={(30:1)}]
	\draw[fermionnoarrow] (-30:1)--(0,0);
	\draw[fermionnoarrow] (30:1)--(0,0);
	\node at (-30:1.2) {$\nu$};
	\node at (30:1.2) {$\nu$};	
\end{scope}
\end{scope}
\end{tikzpicture}
\end{center}
\caption{Diagrams (in the interaction basis) responsible for thermalizing the SM and twin sector. Mass insertions correspond to an insertion of a mixing between a SM and twin neutrino.\label{fig:scattering}}
\end{figure}
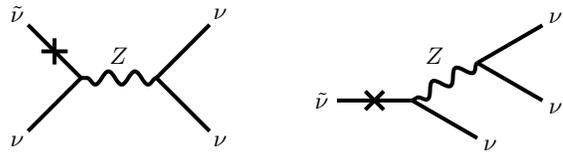

In this section, we explore the consequences of mixing between SM and twin neutrinos on the cosmology of Twin Higgs models. We show that for sufficiently large mixing angles, the neutrino-twin neutrino scattering processes in Fig.~\ref{fig:scattering} may be the last to efficiently transfer energy between the sectors,  thereby lowering the decoupling temperature and potentially reducing the contribution to $N_{\rm eff}$.

Twin neutrino mixing induces interactions, mediated by SM EW gauge bosons, between
the twin neutrinos and the SM leptons:
\bea
\mathcal{L_{{\rm int}(\nu,\tilde\nu)}} &=& \frac{g}{\sqrt{2}}\bar{\ell} \gamma^\mu P_L (c_\theta\nu + s_\theta \tilde{\nu})  W^+_\mu + {\rm h.c.} \no\\
&&+  \frac{g}{2 c_{w}}  (c_\theta\bar{\nu} + s_\theta \bar{\tilde{\nu}}) \gamma^\mu P_L (c_\theta\nu + s_\theta \tilde{\nu})  Z_\mu  \,,
\eea
where $c_\theta = \cos\theta$,  $s_\theta = \sin\theta$, and $\theta$ is the neutrino mixing angle. Throughout this section we will assume that one only twin neutrino mixes with a SM neutrino, while the results are easily generalized to more complicated mixing. 
Energy transfer between the two sectors is most efficient when scattering between sectors involves relativistic, and therefore abundant, particles. Thus for this discussion we will only need to consider the pions, light charged leptons, and neutrinos in each sector.

In order to estimate the decoupling temperature, we follow~\cite{Adshead:2016xxj} and calculate the fractional energy transfer rate
\begin{equation}
\Gamma^E(T) = \begin{dcases}
\frac{n_i n_j \langle \sigma v \Delta E \rangle_{ij \to k\ell}}{\rho} & \rm 2~to~2~scattering \\
\frac{n_i m_i \Gamma_{i \to jk}}{\rho} & \rm  decays~and~inverse~decays
 \end{dcases}	
\end{equation}
where $n_i(n^\prime_i)$ and $\rho_i(\rho^\prime_i)$ are the equilibrium number density and energy density distributions of the particles involved, and $\rho$ is the average density of all the particles involved.
The thermally averaged energy transfer rate is
\bea
 n_i n_j \langle \sigma v \Delta E_i \rangle_{ij \to k\ell} \equiv \int d\Pi_{i} d\Pi_{j} d\Pi_{k} d\Pi_{\ell}   (2\pi )^4\delta^4\!\left(p \right)   \label{sigmavdef}\no\\
 \times f_{i} f_{j} (1 \pm f_k) (1 \pm f_\ell) \overline{\left|M\right|^2}  \Delta E_i \,,
\eea
where $ d \Pi_i \equiv \frac{g_i d^3 p_i}{(2\pi)^32 E_i } $ is the Lorentz invariant phase-space volume. The matrix elements are to be averaged over initial and final degrees of freedom.  To find the decoupling temperature, the thermally averaged rates need to be calculated. This  involves numerically integrating the high dimensional phase-space integrals in Eq.~(\ref{sigmavdef}). When performing the integrals numerically, we follow the techniques given in Appendix A of~\cite{Adshead:2016xxj}.

Thermal decoupling occurs when the energy transfer rate is no longer fast compared to the expansion,
\beq
\Gamma^E(T_d) \simeq H(T_d), \label{thermal}
\eeq
at which point the energy transfer process begins losing to the expansion of the universe and freezes out. We have explicitly solved the full Boltzmann equations and verified that Eq.~(\ref{thermal}) well-approximates the correct decoupling temperature.

The two most significant energy-transfer processes for mixing angles $s_\theta \lesssim 10^{-3}$ are the semi-annihilations $\tilde{\nu} \nu \leftrightarrow \nu \nu$ and decays and inverse decays $\tilde{\nu} \leftrightarrow \nu f \bar{f} $ where $f$ is a light SM fermion. One may expect that twin pion decays and scattering may also be significant, as they are also only suppressed by one mixing insertion. However, these occur via off-shell twin weak bosons and are suppressed by $(v/f)^4$ relative to the twin neutrino processes mediated by SM EW bosons.

\begin{figure}[t!]
\center{
\includegraphics[width=.47\textwidth]{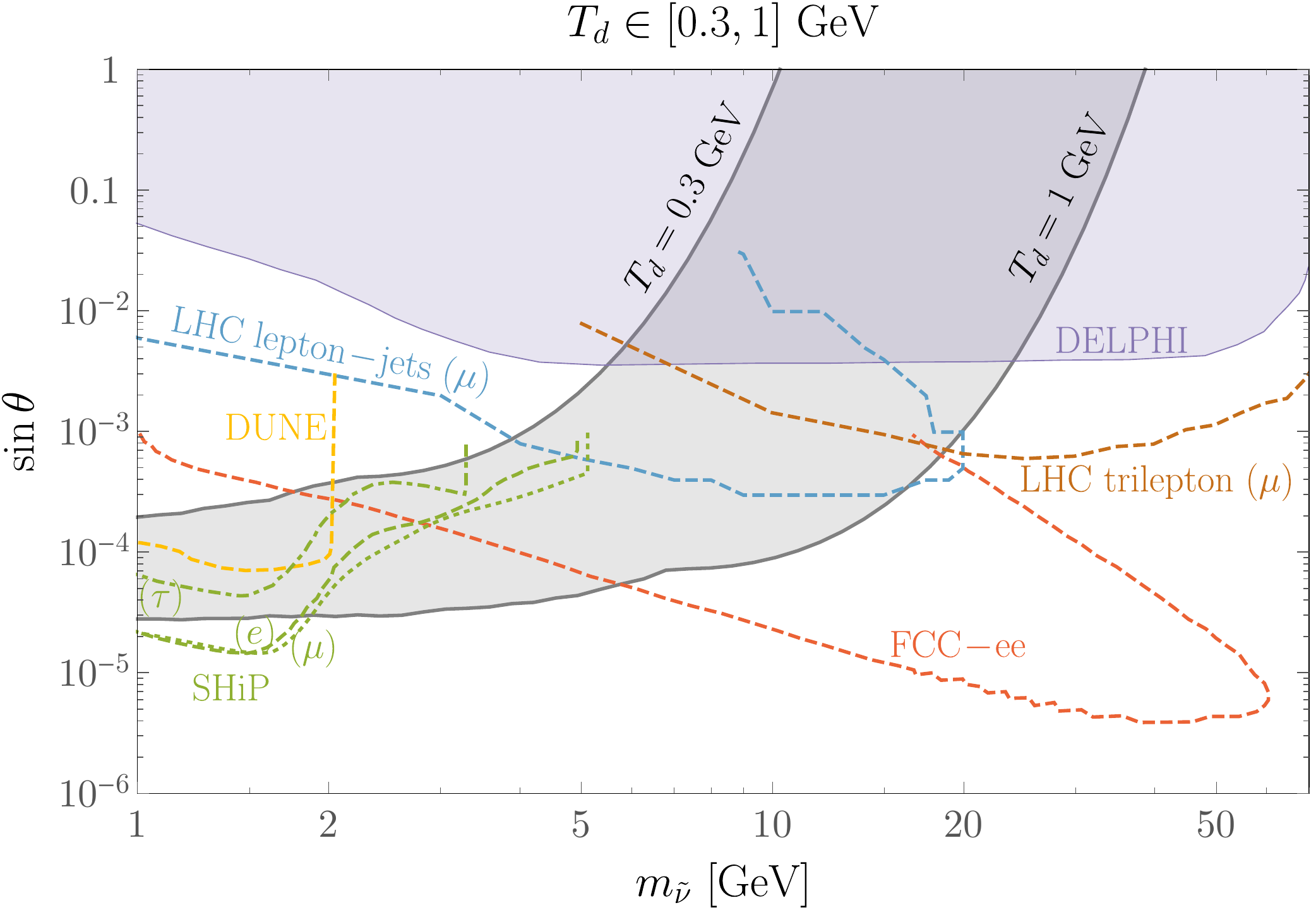}
}
\caption{\label{fig:mvsin} Contours of the cosmologically preferred region $0.3$ GeV $<T_d< 1$ GeV as a function of the twin neutrino mass and mixing, as well as current and projected bounds on sterile neutrinos. The shaded {\color[rgb]{0.53, 0.47, 0.70}\bf purple} region shows the 95\% C.L. limits from DELPHI on sterile neutrinos produced from $Z$-decays at LEP~\cite{Abreu:1996pa}. We also show projected reaches from displaced searches (see next section) as dashed curves. Projections for  SHiP~\cite{Anelli:2015pba} ({\color[rgb]{0.56, 0.69, 0.19}\bf green}), DUNE~\cite{Adams:2013qkq} ({\color[rgb]{1, 0.75, 0}\bf yellow}), and FCC-ee~\cite{Blondel:2014bra,Abada:2014cca} ({\color[rgb]{.92, .38, .2}\bf red}) are taken from \cite{Deppisch:2015qwa}. The LHC reach at $\sqrt{s} = 13$~TeV  with 300~fb$^{-1}$, with searches for lepton jets ({\color[rgb]{0.36,.61,.78}\bf blue})  and trileptons ({\color[rgb]{0.77,.43,.10}\bf brown}) are also shown~\cite{Izaguirre:2015pga}.  The exclusion region and projection curves are only valid if the twin-neutrino cannot decay into light twin particles, for instance, in the FTH (1 generation) scenario.
}
\end{figure}

The energy transfer rate from decays and inverse decays is
\beq
\Gamma_{\tilde{\nu} \leftrightarrow \nu f \bar{f}} = \frac{ \alpha^2 s_\theta^2 m_{\tilde{\nu}}^5 }{48 \pi c_w^4 s_w^4 m_Z^4} \ . \label{gammafff}
\eeq
A fairly good analytic solution for $T_d$ in Eq.~(\ref{thermal}) can be obtained if decoupling happens when some of the particles involved are non-relativistic,  and the phase-space integrals can be performed analytically. In the limit that $T \ll m_{\tilde{\nu}}$, we find for the semi-annihilation process,
\beq
\langle \sigma  v E_{\tilde{\nu}} \rangle_{\tilde{\nu} \nu \to \nu \nu } = \frac{ \pi \alpha^2 s_\theta^2 m_{\tilde{\nu}}^3}{2 c_w^4 s_w^4 m_Z^4}  \ . \label{semiannanalytic}
\eeq
 Using Eqs.~(\ref{gammafff}) and (\ref{semiannanalytic}) in Eqs.~(\ref{sigmavdef}) and~(\ref{thermal}), we find that the process $\tilde{\nu} \nu \leftrightarrow \nu \nu $  thermally decouples  at
 \beq
 \frac{m_{\tilde{\nu} }}{T_d} \simeq 	13.4+\log \left(\frac{s_\theta}{10^{-3} }\right)^2 \left( \frac{m_{\tilde{\nu} }}{10~\rm GeV}\right)^3 \left( \frac{g_{\star s}}{69}  \right)^{\frac{1}{2}}\left( \frac{{T_d}}{13.4} \right)^{\frac{3}{2}}
 \eeq
and   $ \tilde{\nu} \leftrightarrow \nu f \bar{f}  $ decouples at
\beq
 \frac{m_{\tilde{\nu} }}{T_d} \simeq 		17.6+\log \left(\frac{s_\theta}{10^{-3} }\right)^2 \left( \frac{m_{\tilde{\nu} }}{10~\rm GeV}\right)^3 \left( \frac{g_{\star s}}{69}  \right)^{\frac{1}{2}}\left( \frac{{T_d}}{17.6} \right)^{\frac{9}{2}}.
\eeq

 Thus for most of the parameter space, decays and inverse decays decouple later than the semi-annihilation. Ideally, thermal decoupling happens before the QCD phase transition, but below the scales of $\mu^\prime$, $\pi^\prime$ and $\Lambda^\prime$, around $0.3 \lesssim T_d \lesssim 1$~GeV.  In Fig.~\ref{fig:mvsin} we show the contours of this region as a function of $m_{\tilde{\nu}}$ and $\sin\theta$, using the full numerical phase space integrals. As expected, decays and inverse decays provide the last scattering, except for small twin neutrino masses $m_{\tilde{\nu}} \lesssim 5$~GeV, where the width becomes suppressed. In this case semi-annihilations are the last process to decouple, but the twins neutrinos will still be semi-relativistic at decoupling, which slightly reheats the twin sector relative to the SM and enhances $\Delta N_{\text{eff}}$. Optimal values of decoupling occur for  $m_{\tilde{\nu}} \sim 10$~GeV and  $\sin\theta \sim  10^{-4} - 10^{-3}$.

\section{Signatures and Constraints}

The first signal of the Twin Higgs sector may come from a measurement of  $N_{\rm eff}$ at late times. As evident from Fig.~\ref{NeffMTH} the full MTH model is already in tension with data from Planck, while the two-generation model will be probed soon by Stage 3 CMB  experiments. The single generation FTH model, with a massless twin photon or twin neutrino, may not be probed until future Stage 4 CMB experiments. 

In addition to the CMB measurements of $N_{\rm eff}$, we should look for other opportunities to discover the Twin Higgs. The most well-explored direction is to exploit the Higgs portal connecting the SM and twin sectors, leading to displaced Higgs decays for the FTH~\cite{Juknevich:2009gg,Strassler:2006ri,Craig:2015pha,Kang:2008ea,Juknevich:2009ji,Curtin:2015fna,Csaki:2015fba}. We can also look for the exotic states present at the scale of the composite and supersymmetric UV completions~\cite{Goh:2006wj,Cheng:2016uqk,Chang:2006ra,Cheng:2015buv}. 

As we have argued that a sizable mixing between the twin and standard neutrino sectors could be present, we should also be able to probe the twin sector via the neutrino portal. The twin neutrinos can be produced in decays of the $Z$-boson and heavy mesons at high and low energy machines. However, if there are twin particles lighter than the twin neutrinos, the neutrinos will preferentially decay into twin particles which will ultimately be invisible to detection, leading to missing energy signals. If no twin particles are kinematically accessible to the twin neutrino, {\it e.g.} as in a one generation twin model, the twin neutrino will decay back into the visible sector, providing another possible window for detection of TH.

The decay width corresponding to the range of masses and mixing angles which minimize $N_{\rm eff}$ for a one-generation twin model with massless twin photon correspond to macroscopic lifetimes for the twin neutrinos,\footnote{In the minimal FTH, if the twin neutrino is heavy, it must decay---via its mixing with the SM neutrino---before BBN, also motivating macroscopic lifetimes.}
\beq
\tau _{\tilde{\nu }}=0.38~\text{mm} \left(\frac{10^{-3} }{s_{\theta }}\right)^2 \left(\frac{10 \,\text{GeV}}{m_{\tilde{\nu }}}\right)^5\,.
\eeq
This allows the twin neutrinos to be probed at the LHC via displaced vertex searches~\cite{Izaguirre:2015pga} and the proposed MATHUSLA detector~\cite{Chou:2016lxi}; and at fixed target and beam dump experiments such as DUNE~\cite{Adams:2013qkq} and SHiP~\cite{Anelli:2015pba}. Future high energy $e^+e^-$ machines may also probe this parameter space~\cite{Blondel:2014bra,Abada:2014cca}.  The projected reach of these experiments are depicted in Fig.~\ref{fig:mvsin} alongside the cosmologically preferred region.

\section{Neutrino masses and mixing \label{sec:mixing}}
The low energy mass-matrix  involving the SM neutrinos $\nu_R, \nu_L$ and the twin neutrinos $\tilde{\nu}_R,  \tilde{\nu}_L$, is in general an arbitrary 4x4 matrix (where for simplicity we suppress flavor in this and later sections). The form of this mass matrix can be quite complicated in general. In what follows, we draw motivation from the Randall-Sundrum (RS) setup that will be described below and consider two simple benchmark scenarios for neutrino masses and mixings. We find that only the second scenario can lead to large enough mixing for the neutrino-interactions to thermalize the two sectors ($\sin\theta > 10^{-5}$), as we will be shown below.

{\bf Two Seesaws}: The simplest case to  consider is the setup where the only source of $Z_2$ breaking and lepton number violation are the  different seesaw scales. In the RS setup this corresponds to the case when the only source of $Z_2$ breaking and lepton number violation are the right-handed neutrino masses localized on the UV brane. The effective neutrino mass and mixing terms are then
\bea
\mathcal{L} &=& \frac{1}{2} M \nu_R \nu_R + \frac{1}{2}\tilde{M} \tilde{\nu}_R \tilde{\nu}_R  + {M_D} \nu_R \tilde{\nu}_R  \no\\
 && + m_D \nu_R \nu_L  + \tilde{m}_D \tilde{\nu}_R \tilde{\nu}_L + h.c.  \label{neutrino1}
\eea
Assuming $M_D \ll M,\tilde{M}$, the light SM neutrinos and twin neutrinos have a typical see-saw mass set by the scales
\beq
m_{ \nu}\simeq\frac{m_D^2}{M}, ~~~~~~
m_{\tilde \nu}\simeq \frac{\tilde{m}_D^2}{ \tilde{M}},
\label{eq:2seesaws}
\eeq
respectively. Mixing between the twin and SM neutrinos is induced via the UV brane Dirac mass $M_D$ leading to a mixing angle
\beq
\sin\theta  \simeq  \frac{M_D}{\sqrt{M \tilde{M}}} \sqrt{\frac{m_\nu}{m_{\tilde{\nu}}}}\,,
\label{eq:mixingsep}
\eeq
 where $m_\nu $  and $m_{\tilde{\nu}}$ are the light SM and twin neutrino masses. This mixing will be bounded by ${\sin\theta \ll 10^{-6}}$ due to the small neutrino mass ratio, and will generally not be large enough to thermalize the two sectors when the twin neutrino becomes non-relativistic.

{\bf SM seesaw, Dirac twin neutrinos:}
Another interesting limit is the case when only the SM neutrinos are see-sawed, while the twin neutrinos are pseudo-Dirac. In the 5D setup this corresponds to the situation when the singlet twin neutrino is strongly IR-localized and thus cannot feel the UV-localized Majorana masses, while $L-\tilde{L}$ is broken only on the UV brane. The effective neutrino masses and mixings are
\beq
\mathcal{L} = \frac{1}{2} M \nu_R \nu_R  + m_D \nu_R \nu_L  + \tilde{m}_D \tilde{\nu}_R \tilde{\nu}_L +  m \tilde{\nu}_L \nu_L +  h.c. \label{neutrino2}
\eeq
Integrating out $\nu_R$ in this limit,
\begin{equation}
\mathcal{L}_{\rm eff}= -\frac{m_D^2}{2 M}  {\nu_L}^2 + (m  {\nu_L} + \tilde{m}_D  {\tilde{\nu}_R}) {\tilde{\nu}_L} + h.c .
\label{eq:diraceff}
\end{equation}
and ${\tilde{\nu}_L}$ and the linear combination $m {\nu_L} + \tilde{m}_D {\tilde{\nu}_R}$ then form a pseudo Dirac pair with mass $\sim \tilde{m}_D$ for $\tilde{m}_D \gg m$, and the SM neutrino acquires a Majorana mass $m_D^2 / 2 M$. The
mixing angle in this case is
\beq
\sin\theta \simeq \frac{m^2}{2  \tilde{m}_D^2}.
\label{eq:mixing2}
\eeq
Compared the previous two see-saw scenario, this mixing is not limited by the neutrino mass ratio and can be quite large.

\section{Twin neutrinos in a warped UV completion\label{sec:Z2breaking}}

We have shown in the previous sections that the neutrino sector of  Twin Higgs models can have a significant effect on its cosmology. Here we explore the most well-known UV completion of Twin Higgs models based on a warped extra dimension known as the ``holographic composite Twin Higgs" (CTH)~\cite{Geller:2014kta}. This setup can address both the structure of the twin neutrino masses and mixings, as well as the effect of the $Z_2$ breaking necessarily present in the neutrino sector on the Higgs potential.

\subsection{The setup of the holographic CTH}

\begin{figure}[t!]
\center{
\includegraphics[width=.47\textwidth, trim={6cm 6cm 6cm 3cm},clip]{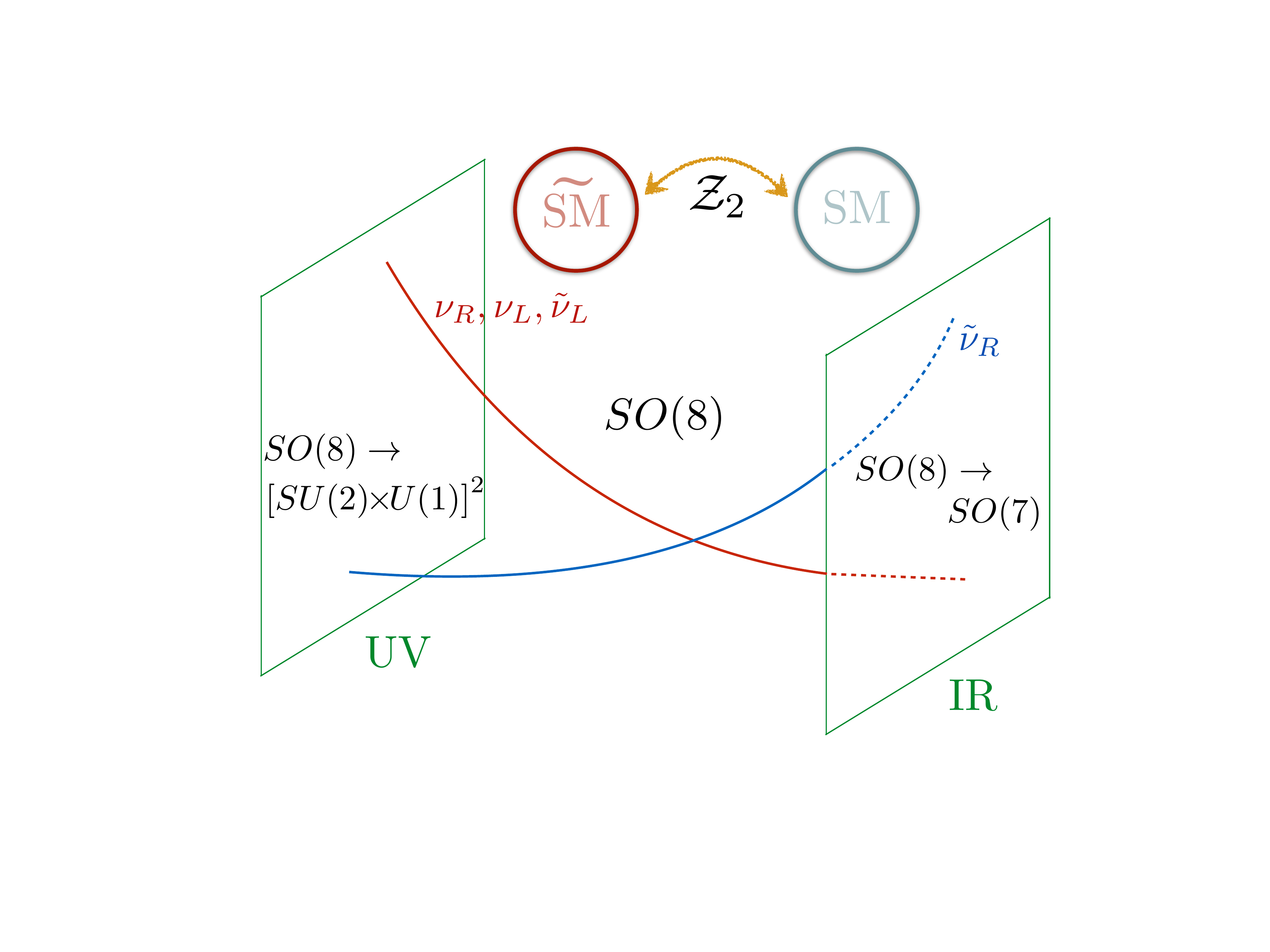}
}
\caption{\label{fig:RS} An illustration of the structure of the neutrino sector in the warped UV completion.
}
\end{figure}

The holographic composite Twin Higgs model is based on a 5D RS~\cite{Randall:1999vf,Randall:1999ee} setup with an AdS$_5$ background metric $ds^2= (R/z)^2 (dx^2-dz^2)$, where $R$ is the AdS curvature, the UV brane is at $z=R$ and the IR brane at $z=R'$. The SO(8) global symmetry required for the Twin Higgs mechanism is incorporated as a bulk gauge symmetry (along with QCD and twin QCD, which play no role in the lepton sector). On the UV brane, boundary conditions for the gauge fields break the bulk gauge symmetry down to the gauge symmetries of the Twin Higgs: SO(8)$\rightarrow$ SU(2)$_L\times$ SU(2)$_L^m \times$ U(1)$_Y\times$ U(1)$_Y^m$. This breaking pattern ensures the correct light gauge boson spectrum\footnote{Although we do not do so, one could eliminate the twin photon from the spectrum by breaking the mirror hypercharge U(1)$_Y^m$ on the UV brane.}. On the IR brane, SO(8) is broken down to SO(7). The gauge generators broken on both branes correspond to broken global symmetries and result in seven Goldstone bosons arising from the $A_5$ component of the corresponding gauge fields. Six of the seven Goldstones are eaten by SM and twin gauge $W$ and $Z$ bosons, and one remains as the physical pseudo-Nambu-Goldstone boson Higgs.

We additionally gauge $(B-L)-(\tilde{B}-\tilde{L})$ in the bulk. This extra gauge symmetry ensures that $L-\tilde{L}$ lepton number is a good symmetry in the bulk and on the IR brane (the CFT interpretation of this statement is that the CFT itself preserves this combination of lepton and twin lepton number). To make sure this does not result in new light degrees of freedom, the symmetry is broken on the UV brane. The UV brane is the only source of $L-\tilde{L}$ violation; thus all Majorana masses arise from UV-localized operators.  An illustration of the structure of the warped UV completion can be seen in Fig.~\ref{fig:RS}.

The two relevant mass scales of the 5D model are $R'$ and $f$. The UV-brane location, $R'$, sets the KK scale ($M_{KK} \approx 2 / R'$), while $f$ is the global symmetry breaking scale for SO(8)$\to$SO(7):
\begin{align}
f = \frac{2}{g_* R'}.
\end{align}
Here $g_*$ is the the dimensionless 5D SO(8) gauge coupling ($g_* = g_5 R^{-1/2}$) which parametrizes the interaction strength of the KK modes and sets the ratio $M_{KK}/f$.\footnote{This $g_*$ should not be confused with the number of relativistic degrees of freedom $g_{\star}$. The strong coupling limit $g_* \sim 4 \pi$ reduces the tension with flavor constraints in the quark sector~\cite{Csaki:2015gfd} and also creates separation between $M_{KK}$ and $f$. At the same time this limit does not increase the Higgs potential tuning because the dominant quantum corrections to the Higgs mass are cut off by the twin top at a scale  $\sim f$. }

\subsection{The neutrino sector of the holographic CTH\label{sec:neutrinosetup}}
\label{sec:neutrinosetup}
The SM and twin SU(2)$_L$ doublet leptons are embedded in two separate vectors ${\bf 8}$ of SO(8)$\supset $SO(4)$\times$SO(4)$^{m} \supset$SU(2)$_L\times$SU(2)$_R\times$SU(2)$^m_L\times$SU(2)$^m_R$ as
  \begin{align}  \label{eq:initialBCs}
    \Psi_8 &= \frac{1}{\sqrt{2}}\begin{bmatrix}
           e_L + \ldots \\
          i e_L + \ldots \\
           \nu_L + \ldots \\
          i\nu_L + \ldots  \\
           \vdots
         \end{bmatrix}, \ \
             \Psi_8^m = \frac{1}{\sqrt{2}} \begin{bmatrix}
             \vdots \\
   		\tilde{e}_L+\ldots\\
           i \tilde{e}_L+\ldots\\
           \tilde{\nu}_L+\ldots \\
         i\tilde{\nu}_L+\ldots \\
         \end{bmatrix}
  \end{align}
 and the right handed neutrinos are introduced as SO(8) singlet fermions
\begin{equation}
\Psi_1 = \nu_R , \ \Psi_1^m = \tilde{\nu}_R.
\end{equation}
Here and throughout we neglect flavor indices. One  can understand the discussion below as pertaining to one twin neutrino generation, while generalizing to multiple generations is straightforward.

The symmetry breaking patterns on the branes, along with the $L-\tilde{L}$ symmetry, will determine which low-energy mass terms exist for the would-be zero modes. Since UV-localized mass terms are the only source of $L-\tilde{L}$ breaking, the effective zero-mode low-energy Lagrangian, after integrating out the KK modes, is
\bea
\mathcal{L} &=& \frac{1}{2} M \nu_R \nu_R + \frac{1}{2}\tilde{M} \tilde{\nu}_R \tilde{\nu}_R  + {M_D} \nu_R \tilde{\nu}_R  \no\\
 && + m_D \nu_R \nu_L  + \tilde{m}_D \tilde{\nu}_R \tilde{\nu}_L+  m \tilde{\nu}_L \nu_L  + h.c.  \label{Left}
\eea

Depending on the size of the mass-terms in Eq.~(\ref{Left}), this matches onto either Eq.~(\ref{neutrino1}) or Eq.~(\ref{neutrino2}). In the limit where mixing terms between the SM and twin sectors are negligible ($M_D$, $m \rightarrow 0$), and assuming all neutrino fields are UV-localized, the Majorana mass terms arising from the UV brane ($M$, $\tilde{M}$) will be large, and we arrive at the two seesaws of Eq.~(\ref{eq:2seesaws}). We note, however, that this situation is tuned: $M_D$ is naturally the same size as $M$ and $\tilde{M}$.\footnote{The $m$ term can be forbidden naturally by gauging $B-L$ and $\tilde{B}-\tilde{L}$ separately in the bulk.} If instead, the right-handed twin neutrino is very IR-localized, with all other neutrino fields kept UV-localized, $\tilde{M}$ and $M_D$ are exponentially suppressed by $R / R^\prime$ (see Eq.~(\ref{rsmasses})), and we match onto Eq.~(\ref{neutrino2}).

\subsection{$Z_2$-breaking effects on the Higgs potential}
Finally, we explore the consequences of breaking the Twin Higgs $Z_2$ symmetry in the neutrino sector by a large difference between the SM and twin seesaw mass scales.  The $Z_2$ breaking in the neutrino sector reintroduces quadratic sensitivity to the cutoff scale into the Higgs mass term from one-loop diagrams involving the would-be zero mode neutrinos and  KK modes. Without the warped UV completion, one might naively expect the contributions from the $Z_2$ breaking in the neutrino sector to be large---the right-handed neutrino mass provides a large scale which could potentially feed into the Higgs potential. However, in the RS UV completion, the $Z_2$ breaking reintroduces quadratic sensitivity to $M_{KK}$ but not to the seesaw scales themselves.

If the $Z_2$ symmetry is badly broken in the low-lying KK modes, one naively expects the contribution to the Higgs potential to be
\begin{equation}
	\delta m_h^2 \sim \frac{y^2 g_*^2}{ 4 \pi^2} M_{KK}^2\,,
	\label{eq::muterm_naive_KKmode}
\end{equation}
where $y$ is an order one factor arising from the overlap of the KK modes and Higgs. However, this overestimates the corrections to the Higgs mass since it does not take into account the collective effect of all the KK modes---at large momenta, where the KK modes become important, the exponential suppression becomes important. Therefore, we must calculate the Higgs potential in the 5D theory, taking into account all of the KK modes.

We present the detailed calculation  the of 1-loop contributions to the Higgs potential in the Appendix. Since the full formulae for the corrections to the potential in terms of the RS parameters are lengthy and not very illuminating, we present here the approximate expression instead: We will be interested in the case where the twin right-handed neutrino is strongly IR localized and left-handed neutrino is UV localized,  as this is the case that can generate large mixing between the SM and twin neutrinos. In the limit that $c_1^m \gg -0.5$, the contribution to the Higgs mass is dominated by the twin sector and given approximately by
\bea
\delta m_h^2 \simeq
\frac{ g_*^2}{ 4 \pi^2} M_{KK}^2 \frac{\left(2c_8^m - 1\right)}{4} \left(\frac{R}{R'} \right)^{2c_8^m -1} \Bigg[\left( \frac{128}{3 e^4}\right) \times \nonumber \\
  I_{\hf -c_8^m}(x_1) \left( I_{\hf+c_8^m}(x_1) +\frac{I_{\hf+c_1^m}(x_1) I_{\hf -c_8^m}(x_1)}{|m_\nu|^2 I_{\hf-c_1^m}(x_1) }\right) \Bigg]^{-1} \label{deltamhapprox}
\eea
where $x_1 = M_{KK} R^\prime$ and the numerical value of the term in $[\ldots]^{-1}$ is $\mathcal{O}(0.1)$. This formula was obtained by approximating the form factors in Eq.~(\ref{eq:formfactor}) by their high energy behavior, $e^{-4 p / M_{KK}}$. The analogous formula with $c_8^m (c_1^m) \rightarrow c_8 (c_1)$ also holds for the subdominant SM contribution. Compared to the naive estimate, there is a warping down by a factor of $ \left( 2 c_8^m - 1\right)/4 \left({R}/{R'} \right)^{2c_8^m -1}$. This is the same suppression factor that appears in the size of the mixing angle, $\sin\theta$ in Eq.~(\ref{rsmixing}), and twin-neutrino mass, $\tilde{m}_{D}$ in Eq.~(\ref{rsmasses}) when the twin neutrino is strongly IR localized. The numerical value of the contribution to the Higgs mass for a twin neutrino with mass $m_{\tilde{\nu}} = 10$~GeV and mixing $\sin\theta = 10^{-4}$ is small for most parameters. 
However Eq.~(\ref{deltamhapprox}) is an approximate expression and breaks down for $c_8^m $ very close to 1/2, i.e., where the twin neutrino becomes heavy. There are points in parameter space where a single twin neutrino can have sizable mixing, and the KK modes contribute significantly to the $Z_2$-breaking Higgs mass term.

In addition, the different generations of twin neutrinos are not necessarily expected to be degenerate, and a heavy twin neutrino can give a large contribution to $\delta m_h^2$ if $m_{\tilde{\nu}} \sim 200$~GeV. 
For the case of a heavy twin-neutrino, the contribution to $\delta m_h^2 $ can be well approximated by the would-be zero mode alone.
Calculating the standard one-loop contribution to the Higgs potential from the Yukawa coupling $ \frac{m_{\tilde{\nu}}}{f} h \tilde{\nu}_L \tilde{\nu}_R  $, we find
\bea
	\delta m_h^2 &\simeq& \frac{1}{ 4 \pi^2}
	\left( \frac{m_{\tilde{\nu}}}{f} \right)^2 M_{KK}^2  \\
	&\simeq& \left(260~{\rm GeV}\right)^2 \left(\frac{m_{\tilde{\nu}}}{200~ \rm GeV}\right)^2 \left(\frac{M_{KK}^2}{10~ \rm TeV}\right)^2 \left(\frac{5}{f/v} \right)^2\nonumber\,.
	\label{eq::muterm_naive}
\eea

\vspace{.5cm}
\begin{acknowledgments}
{\em Acknowledgments ---}
We thank  Kaustubh Agashe, Marco Farina, Michael Geller, Yonit Hochberg, Sungwoo Hong, Jessie Shelton, Ofri Telem and Yuhsin Tsai for useful discussions.  C.C., E.K. and S.L. are supported in part by the NSF grant PHY-1316222. EK is supported by a Hans Bethe Postdoctoral Fellowship at Cornell. This work was initiated at the Aspen Center for Physics, which is supported by NSF grant PHY-1066293.
\end{acknowledgments}

\newpage
\begin{appendix}
\vspace{.5cm}
\begin{center}
{\bf APPENDIX}
\end{center}

	\section{RS setup \label{app:RS}}

Here we present more details of the embeddings of the leptons into the warped model.
All fermions are introduced as 5D bulk fields, corresponding to 4-component Dirac spinors, $\textit{i.e.}$ $\Psi = (\chi, \bar{\psi})$, where both $\chi$ and $\psi$ are left-handed 2-component Weyl spinors. The fields explicitly depicted in Eq.~(\ref{eq:initialBCs}) will contain zero modes in the $\chi$ components of the bulk fields (SM and twin left-handed leptons). Their $\chi$ components are assumed to have $(+,+)$ boundary conditions at the UV and IR branes to ensure the existence of a zero mode in $\chi$, while the fields suppressed in Eq.~(\ref{eq:initialBCs}) are assumed to have $(-,+)$ boundary conditions to avoid extra light fermions in the spectrum. Fields containing right-handed zero modes ($\nu_R$, $\tilde{\nu}_R$, $e_R$, $\tilde{e}_R$) are assumed to have $(-,-)$ boundary conditions for the $\chi$-component, in order to allow for a zero mode in the $\psi$ component of the corresponding bulk field.

The vector of SO(8) contains two SU(2)$_L$ doublets, $q_{\pm \pm}$, where the first and second subscript denotes the field's $T^3_L$ and $T^3_R$ quantum numbers, respectively, and two SU(2)$_L^m$ doublets $p_{\pm \pm}$,
\[ \begin{array}{r}
    \Psi_8 = \frac{1}{\sqrt{2}} \left(
           q_{++} \!+\! q_{--} ,
           i q_{++} \!-\!  iq_{--},
           q_{+-} \!+\!  q_{-+},
			iq_{+-} \!-\! iq_{-+}, \right. \\
		\left.	 p_{++} \!+\!  p_{--} ,
           i p_{++} \!-\! ip_{--},
           p_{+-} \!+\!  p_{-+},
           ip_{+-} \!-\! ip_{-+}\right).
  \end{array}\]
  We identify hypercharge as $Y=T^3_R$ and twin hypercharge as $Y^m = T^{3m}_R$, so $q_{+-} $ has the quantum numbers of $ \nu_L$ and $q_{--} $ has the quantum numbers of $ e_L$. For the twins states, we should identify $p_{+-}$ with $\tilde{\nu}_L$ and $p_{--}$ with $\tilde{e}_L$. In addition to the fields defined earlier, right-handed electrons are embedded into the antisymmetric $\bf{28}$ representation of SO(8), $
\Psi_{28},\ \Psi_{28}^m$.

Each 5D field has a bulk Dirac mass $\frac{c}{R} \bar{\Psi} \Psi$, where the dimensionless parameters   $c_8$, $c_8^m$, $c_1$, $c_1^m$, $c_{28}$, $c_{28}^m$ control the localization properties of the zero modes. We consider parameter space where all lepton fields are UV localized (corresponding to elementary leptons in the 4D language), except the singlet twin neutrino which we allow to be either UV localized ($c_1^m < -\hf $) or IR localized ($c_1^m > -\hf$), the latter case corresponding to composite right-handed twin neutrinos.

The symmetry breaking patterns allow mass terms on the UV and IR branes, which will determine the masses and mixings among the light neutrinos.  On the UV brane, the most general renormalizable Lagrangian allowed by the gauge symmetries and the boundary conditions includes the mass terms
\begin{equation}
\mathcal{L}_{UV} = \hf M_{\nu_R} \psi_{\nu_R} \psi_{\nu_R} + \hf M_{\tilde{\nu}_R} \psi_{\tilde{\nu}_R} \psi_{\tilde{\nu}_R}
+ M' \psi_{\nu_R} \psi_{\tilde{\nu}_R} + h.c.
\label{eq:LUV}
\end{equation}
where the brane mass parameters $M_{\nu_R}, M_{\tilde{\nu}_R}$ and $M'$ are dimensionless and are generically $3\times 3$ matrices if we consider the full 3-generation twin sector. We note that both lepton number and twin lepton number, as well as the $Z_2$ symmetry relating the two sectors is broken on the UV brane. In particular, $Z_2$ is broken on the UV brane if $M_{\nu_R} \neq M_{\tilde{\nu}_R}$. The Majorana mass terms $M_{\nu_R}$ and $M_{\tilde{\nu}_R}$ for the singlet neutrinos lead to the warped seesaw mechanism~\cite{Huber:2003sf, Csaki:2003sh, Agashe:2015izu}. The CFT interpretation of the warped seesaw mechanism (see~\cite{Agashe:2015izu}) is that while the CFT itself is lepton-number preserving, the elementary sector breaks lepton number at a high scale.

Using the interpretation that the CFT is $Z_2$ preserving, we require that the IR brane localized mass terms are $Z_2$ invariant. On the IR brane SO(8) is broken to SO(7), under which $\Psi_8$ decomposes as  $\Psi_8^{\bf 7} + \Psi_8^{\bf 1}$. We can then write the following SO(7) invariant IR brane localized mass terms
\begin{align}
\mathcal{L}_{IR} &= - \left(\frac{R}{R'}\right)^4 [ m_{\nu} (\chi^1_8 \psi_{\nu_R} + \chi^{1m}_8 \psi_{\tilde{\nu}_R})
+m^\prime \chi^1_8 \chi^{1m}_8 \nonumber
\\ &~~~~+ m_e (\chi_8^7 \chi_{28}^7
+ \chi_8^{7m} \chi_{28}^{7m}) ]
 + h.c. \label{eq:LIR}
\end{align}
which will be responsible for the Dirac masses of the leptons. The effect of the $\chi^1_8 \psi_{\nu_R}$ operator is to modify the boundary conditions on the IR brane such that the right-handed singlet neutrino zero mode in $\psi_{\nu_R}$ is partially rotated into $\chi^1_8$. Once $A_5$ acquires a vev (corresponding to EWSB), the two would-be zero modes in $\chi_8$ acquire a Dirac mass from $\overline{\Psi}_8 \langle A_5 \rangle \Psi_8$. Similarly $\chi^1_8 \chi^{1m}_8$ rotates the zero mode in $\chi_{\tilde{\nu}_L}$ (living in $\chi^{1m}_8$) into $\chi^{1}_8$ and leads to a Dirac mass between $\chi_{\nu_L}$ and $\chi_{\tilde{\nu}_L}$ after EWSB.

\section{Mass spectrum \label{app:mass}}

In this Appendix we describe the mass terms for the would-be zero mode leptons to lowest order in $R/R'$ and $v/f$. We first observe that all KK states can be integrated out, leaving only the effective Lagrangian for the would-be zero modes
 \bea
 \label{eq:EFTzeromodes}
\mathcal{L}_{\rm eff} &=& \hf M\psi_{\nu_R}^{(0)} \psi_{\nu_R}^{(0)}
+\hf \tilde{M}\psi_{\tilde{\nu}_R}^{(0)} \psi_{\tilde{\nu}_R}^{(0)}
+ M_D\psi_{\nu_R}^{(0)} \psi_{\tilde{\nu}_R}^{(0)} \\
&&+ m_D \chi_{\nu_L}^{(0)} \psi_{\nu_R}^{(0)}
+ \tilde{m}_D \chi_{\tilde{\nu}_L}^{(0)} \psi_{\tilde{\nu}_R}^{(0)} + m \chi_{\nu_L}^{(0)} \chi_{\tilde{\nu}_L}^{(0)}
+ h.c. \nonumber
\eea
which matches onto Eq.~(\ref{Left}).
Majorana mass terms $\chi_{\nu_L}^{(0)} \chi_{\nu_L}^{(0)}$ and $\chi_{\tilde{\nu}_L}^{(0)} \chi_{\tilde{\nu}_L}^{(0)}$ do not  appear even after EWSB since the Higgs sector preserves $L-\tilde{L}$ (a result of the fact that $B-L-\tilde{B}+\tilde{L}$ is a gauge symmetry in the bulk and IR brane).

Expressing the (dimensional) mass terms for the would-be zero mode neutrinos in terms of the RS parameters, we find  
\begin{equation}\def\arraystretch{1.5}\begin{array}{cl}
m_D &\simeq \frac{g_* v m_\nu}{2} a_{c_1,-c_8}\left(  \frac{R}{R'} \right)^{c_{8}-c_{1}-1} \\
\tilde{m}_D &\simeq
\frac{g_* f m_\nu}{2} \times \begin{cases}
a_{c_1^m,-c_8^m} \left(  \frac{R}{R'} \right)^{c_{8}^m-c_1^m-1}  & c_1^m < -\hf \\
i a_{c_1^m,-c_8^m}  \left(  \frac{R}{R'} \right)^{c_{8}^m-\hf} & c_1^m > -\hf	
\end{cases}
\\
m &\simeq \frac{g_* v m^\prime}{2} a_{-c_8,-c_8^m}  \left( \frac{R}{R'} \right)^{c_8 + c_8^m -1}
\\
M &\simeq
-(1+2c_1)\frac{M_{\nu_R}}{R}
\\
\tilde{M} &\simeq
\frac{ M_{\tilde{\nu}_R}}{R} \times \begin{cases}
-(1+2c_1^m) & c_1^m < -\hf \\
(1+2c_1^m) \left( \frac{R}{R'} \right)^{1+2c_1^m} & c_1^m > -\hf
 \end{cases}
 \\
M_D &\simeq
\frac{M'}{R} \times \begin{cases}
\sqrt{2} a_{c_1, c_1^m} & c_1^m < -\hf \\
\sqrt{2} i a_{c_1, c_1^m} \left( \frac{R}{R'} \right)^{\hf+c_1^m}	& c_1^m > -\hf
\end{cases}
\end{array}\label{rsmasses}\end{equation}
where $a_{x,y}= \sqrt{(1+2 x)(1+2y)/2}$.
Note that keeping terms higher order in $v/f$ results in the replacement $g_* v \rightarrow g_* f \sin(\frac{v}{f})$ and $g_* f \rightarrow g_* f \cos(\frac{v}{f})$, as expected from a pseudo Goldstone (Twin) Higgs. In the absence of mixing, the neutrino mass is given by $m_D^2 / 2 M$. These results agree with the exact result obtained from the lowest zero of the spectral functions, which takes into account the mixing of the KK modes and zero modes within each SO(8) multiplet.  In terms of the RS parameters, the mixing angle in Eq.~(\ref{eq:mixing2}) is
\beq
\sin\theta \simeq \frac{\left(2 c_8^m-1\right)}{4 }\left(\frac{R}{R'}\right)^{2 c_8^m-1}  \frac{1}{c_1^m}\left(\frac{v}{f}\right)^2 \left(\frac{m'}{m_{\nu }}\right)^2 . \label{rsmixing}
\eeq

The electrons (and muons/taus), acquire the following mass terms after EWSB,
\begin{equation}
m_{e} \simeq
\frac{g_* v m_e}{2} \begin{cases}
a_{c_{28},-c_{8}}\left(  \frac{R}{R'} \right)^{c_{8}-c_{28}-1}  & c_{28} < -\hf \\
i a_{c_{28},-c_{8}}\left(  \frac{R}{R'} \right)^{c_{8}-\hf} & c_{28} > -\hf	
\end{cases}
\end{equation}
and similarly for $m_{\tilde{e}} $.

	\section{Coleman-Weinberg potential for Majorana KK spectrum\label{app:CWpotential}}

The calculation presented in the main text of the $Z_2$-breaking effects in the Higgs potential from the neutrino sector is utilizing the full expression of the Coleman-Weinberg potential. However due to the appearance of the Majorana masses for the right handed neutrinos the standard techniques (which assume Dirac fermions) for evaluating the CW potential for KK theories have to be augmented. Here we explain how to deal with a Majorana KK spectrum in general, and present the actual calculation in the next Appendix. 

The general expression of the Coleman-Weinberg potential for KK theories takes the form~\cite{Falkowski:2006vi}
\begin{equation}
V = (-1)^F \frac{N}{2} \sum_n \int \frac{d^4p}{(2\pi)^4} \log(p^2 + m_n^2)
\label{eq:CWpot}
\end{equation}
where $N$ is the number of DOFs at each level of the KK tower (3 for a gauge boson, 4 for a Dirac fermion) and $m_n(v)$ are the KK masses which depend on the Higgs vev. The sum is usually turned into a contour integral in the the complex $m$-plane, resulting in an integral over the spectral function $\rho(m)$, whose zeros encode the KK spectrum. The spectral function is determined by solving the equations of motion (EOM) and applying the boundary conditions to obtain a quantization condition. 

The function $\frac{1}{\rho(m)}$ has simple poles along the real axis corresponding to the KK spectrum, so the the sum over the KK masses can be performed via a contour integral in the complex $m$-plane using zeta function regularization techniques~\cite{Oda:2004rm,Goldberger:2000dv}. For Dirac KK modes, $\rho(m) \equiv \rho(m^2)$ (as consequence of the exact degeneracy of the left and right 2-component spinors which make up each Dirac KK state). Thus for the Dirac case one can perform the KK sum by summing over the positive masses in Eq.~(\ref{eq:CWpot}) and setting $N=4$ to account for the DOFs associated with both chiralities. This is equivalent to a contour integral which only encloses the the positive zeros of the spectral function in the complex $m$ plane. However, in the presence of Majorana mass terms on the UV brane, the KK spectrum becomes pseudo Dirac, and the zeros of $\rho(m)$ at $m \simeq \pm m_0$ do not pair up exactly (and as a consequence $\rho(m)$ is a function of $m$, not $m^2$ as in the Dirac case). Therefore, one must also include the negative mass solutions in the sum in Eq.~(\ref{eq:CWpot}), or equivalently the zeros of $\rho(m)$ along the $\text{Re} (m) < 0$ axis must also be enclosed by a contour.

After throwing away an $h$-independent contribution to the cosmological constant and performing two contour integrals, one to enclose the poles on the $\text{Re}(m)>0$ axis and one to enclose the poles on the $\text{Re}(m) < 0$ axis, we arrive at an integral along the imaginary $m$ axis ($m = i p$)
\begin{equation}
V= (-1)^F \frac{2N}{(4\pi)^2} \int_0^{\infty} dp \, p^{3} {\rm Re} \log[\rho(ip)]
\label{eq:result}
\end{equation}
where $\rho(i p)$ is generally complex since $\rho$ is a function of $m$ rather than $m^2$ and $N=2$ for Majorana KK modes.

\section{Coleman-Weinberg potential for the warped twin seesaw}

We first present the spectral functions $\rho_\nu(m)$ ($\rho_\nu^m(m)$), which encode the exact masses of the SM (twin) neutrino KK spectrum. In the presence of $\nu$-$\tilde{\nu}$ mixing, there is only one spectral function incorporating both the SM and twin neutrino KK towers. However for the purpose of calculating the CW potential we can set the mixing to zero ($M^\prime \rightarrow 0$, $m^\prime \rightarrow 0$) since the effects from the mixing are expected to be small. The two spectral functions are parameterized by two form factors:

\begin{align}
& \rho_\nu = 1 + f_\nu \sin^2 \left({\frac{h}{f}} \right) \\
& \rho_{\nu}^m = 1 + f_{\nu}^m \cos^2 \left({\frac{h}{f}} \right).
\end{align}
The $\sin \left(h/f\right)$ terms are generated by the SM neutrino sector, while the pieces with $\cos\left(h/f\right)$ are generated by twin neutrinos.

First, we look for separable solutions to the bulk EOM, assuming the ansatz
\begin{equation}
\chi = \sum_n g_n(z) \chi_n(x)\ ~~\text{and}\ ~~ \bar{\psi} = \sum_n f_n(z) \bar{\psi}_n(x) .\label{eq::fermion_ansatz}
\end{equation}
In the presence of $\langle A_5 \rangle$, the bulk EOMs are coupled. We can, however, perform a field redefinition (which resembles a gauge transformation within the bulk) to remove $\langle A_5 \rangle$ from the EOMs~\cite{Falkowski:2007kd,Csaki:2008zd}. The transformation that does this is the Wilson line $\Omega(z) = e^{- i g_5 \int dz \langle A_5 \rangle}$. 
We define hatted fields which do not depend on $\langle A_5 \rangle$
\begin{align}
f(z, \langle A_5 \rangle) &= \Omega(z) \hat{f}(z,0)	\nonumber \\
g(z, \langle A_5 \rangle) &= \Omega(z) \hat{g}(z,0)	.
\label{eq::redef}
\end{align}

The differential equations for $\hat{f},\hat{g}$ are now the standard bulk EOMs without the $A_5$ vev, which have solutions in terms of Bessel functions $J_\nu$ and $J_{-\nu}$. It is more convenient to work in the basis of the warped analog of flat space sines and cosines~\cite{Falkowski:2007kd} satisfying $C_c(R) = 1$, $S_c(0)=0$, $C_c'(R)=0$, and $S_c'(R)=m$ where $c$ is the bulk mass parameter, leading to a simpler quantization condition. The explicit expressions for these $C_c, S_c$  functions are 
\begin{align}
S_c (z) &= \frac{ \pi m R}{2} \left( \frac{R}{z} \right)^{-c-\hf} \times \\
& \left(
J_{c+\hf}(m R) Y_{c+\hf}(m z)
- Y_{c+\hf}(m R) J_{c+\hf}(m z) \right) \nonumber \\
C_c (z) &= \frac{\pi m R}{2} \left( \frac{R}{z} \right)^{-c-\hf} \times \\
&\left(
Y_{c-\hf}(m R) J_{c+\hf}(m z)
- J_{c-\hf}(m R) Y_{c+\hf}(m z) \right) \nonumber
\end{align}
One needs to multiply by $z^{2-c}$ in order for these functions to satisfy the bulk EOMs.
 
The boundary conditions are modified by the field redefinition in Eq.~(\ref{eq::redef}). The UV boundary conditions are unchanged since $\Omega(z=R)=0$. On the IR brane however, we must apply Wilson line transformation $\Omega(R^\prime)$ relating $f$ to $\hat{f}$ at $z=R^\prime$, which was worked out in~\cite{Csaki:2015gfd}. Applying the IR boundary conditions to $\Omega \Psi_8$ and $\Omega \Psi_8^m$ produces two 4 by 4 systems of equations involving the neutrino and twin neutrino wave functions. There exists a solution if and only if the determinant of the coefficient matrix is 0. Thus the determinant of each IR boundary condition matrix gives the spectral function of the theory, resulting in 
\begin{eqnarray}
&&\hspace*{-0.3cm}f_{\nu}(p^2) = \nonumber \\ &&\hspace*{-0.3cm}- \hf \frac{|\hat{m}_{\nu}|^2 ( C_{-1}+M_{\nu_R}S_{-1} )}{
S_{-8}[C_{-8}(S_{1}-M_{\nu_R}C_{1}) +|\hat{m}_{\nu}|^2 S_{8}(C_{-1}+M_{\nu_R}S_{-1})]
} \nonumber \\ \label{eq:formfactor}
\end{eqnarray}
where $\hat{m}_{\nu} = m_{\nu} \left( \frac{R}{R'} \right)^{c_8-c_1}$ and the twin form factor has the same form with $c_8 \rightarrow c_8^m$, $c_1\rightarrow c_1^m$, $M_{\nu_R} \rightarrow M_{\tilde{\nu}_R}$. Expanding in small parameters $R / R'$ and $v/f$ to leading order reproduces the masses of the lightest modes obtained from taking into account only the would-be zero modes and neglecting the KK modes. In the limit that the Majorana masses $M_{\nu_R}$, $M_{\tilde{\nu}_R} \rightarrow 0$, the form factors reproduces the the top sector form factors in the CTH~\cite{Csaki:2015gfd}. 

The neutrino sector contribution to the Coleman-Weinberg potential for the Higgs can be written in terms of the spectral functions as
\begin{equation}
V_{\rm eff}(h) = -\frac{4}{(4 \pi)^2} \int_0^{\infty} dp p^3 {\rm Re} \log[\rho_\nu (ip) \rho_{\nu}^m (ip) ].
\end{equation}
In principle, the potential can be calculated numerically by inserting the form of the spectral functions. However, we obtain an analytical approximation for the potential by expanding in powers of $\sin^2(\frac{h}{f})$.

If there were no $Z_2$ breaking, the Twin Higgs $Z_2$ would guarantee that the quadratically divergent pieces cancel, see~\cite{Csaki:2015gfd}. However, we are interested in breaking the $Z_2$ in the neutrino sector. The Higgs mass correction can be found by differentiating with respect to $x = \sin^2(\frac{h}{f})$ and is given by

\begin{align}
\delta m_h^2 &= \frac{1}{f^2} \frac{\del V}{ \del x} (0) \\
&\simeq 
\frac{-4}{(4 \pi)^2 f^2} \int_0^{\infty} dp p^3 \ 
\left( {\rm Re} (f_\nu) - {\rm Re} (f_{\nu}^m) \right)+ \mathcal{O}(f_\nu^2)
\label{eq:potential} \nonumber
\end{align}
where terms quadratic and higher in the form factors have been neglected since the form factors are exponentially suppressed for $p \gtrsim M_{KK}$. 

Without $Z_2$ breaking, ${\rm Re} f_\nu - {\rm Re} f_{\nu}^m = 0$ and the largest possible contribution to the potential from the neutrino sector vanishes up to terms higher order in the form factors. However, if we allow the $Z_2$ to be broken, ${\rm Re} f_\nu - {\rm Re} f_{\nu m} \neq 0$, and we can obtain potentially large corrections to the Higgs mass term.

\end{appendix}

\bibliographystyle{h-physrev}
\bibliography{bib}

\end{document}